\documentclass[12pt]{article}
\usepackage{amsmath}
\usepackage{amsfonts}
\usepackage{amscd}
\usepackage{amsthm}
\usepackage{amssymb}
\usepackage{latexsym}
\newcommand{\nc}{\newcommand*}

\nc{\reff}[1]{(\ref{#1})} 
\newcommand{\bit}{\begin{itemize}}
\newcommand{\eit}{\end{itemize}}
\newcommand{\rd}{\rm{d}}
\newcommand{\ri}{\rm{i}}
\usepackage[cp1251]{inputenc}
\inputencoding{cp1251}
\usepackage[T2B]{fontenc}
\usepackage[russian,english]{babel}
\captionsrussian
\begin{document}
{}
\medskip

\centerline {\Large\bf On calculation of generating functions of }
\bigskip
\centerline {\Large\bf generalized Chebyshev polynomials in several variables
\footnote {The work is supported by RFBR grant \No 15-01-03148-а and partially (PPK)
 by grant \No 14-01-00341 and the programme ``Mathematical problems of nonlinear dynamics'' of RAS.}}

\vspace{1cm}
\begin{center}
E.V. Damaskinsky,\, $^*$\,\footnote{evd@pdmi.ras.ru}\qquad
P.P. Kulish,\, $^\dag$\,\footnote{kulish@pdmi.ras.ru}\qquad
M.A. Sokolov\, $^\ddag$\,\footnote{masokolov@gmail.com}
\bigskip

$^*$\,  Military institute (technical engineering)

\medskip

$^\dag$\,  St. Petersburg Department
of Steklov Institute of Mathematics, \\
Russian Academy of Science
\medskip

$^\ddag$\, St. Petersburg  State Polytechnical University

\end{center}

\vspace{1cm}

\centerline{Abstract}
\bigskip

\begin{quote}
We propose a new method of calculation of generating functions of Chebyshev polynomials in several variables associated with root systems of simple Lie algebras. We obtain the generating functions of the polynomials in two variables  corresponding to the Lie algebras $C_2$ and $G_2$.
\end{quote}

\newpage

\section{Introduction}

In this paper we propose a new method of calculation of generating functions of the generalized Chebyshev polynomials associated with root systems of simple Lie algebras. The method is illustrated by calculations of the generating functions of the
two-variables Chebyshev polynomials associated with the simple Lie algebras $A_2$, $C_2$ and $G_2$.

The Chebyshev polynomials in several variables are a natural generalization of the classical polynomials in one variable \cite{S}. They are used in different areas of mathematics (for example, in discrete analysis,
in approximation theory \cite{RM-K}, in linear algebra \cite{SS}, \cite{AS}, in representation theory \cite{KLP} - \cite{L})
and in physics \cite{GR} - \cite{BD1}.

The classical Chebyshev polynomials of the first kind $T_n(x)$ is defined by the formula
\begin{equation}
\label{ic-0}
T_n(x)=T_n(\cos{\phi})=\cos{n\phi},\quad n=0,1,..\,.
\end{equation}
They satisfy the following three-term recurrence relation
\begin{equation}
\label{ic-1}
T_{n+1}(x)=2xT_n(x)-T_{n-1}(x),\quad n\ge 1,
\end{equation}
with the initial conditions
\begin{equation}\label{ic-2}
T_0(x)=1,\quad T_1(x)=x.
\end{equation}

The simplest $d$-dimensional generalization of the Chebyshev polynomials is reduced to the product of one-dimensional polynomials $T_{n_i}(x_i)$
$$
T_{\bf n}({\bf x})=\prod_{i=1}^dT_{n_i}(x_i),\quad {\bf n} = (n_1,...,n_d),\quad {\bf x} = (x_1,...,x_d).
$$
However, much more interesting multidimensional analogues of the polynomials (\ref{ic-0}) which are associated with roots systems of simple Lie algebras \cite{K1} - \cite{LU}. Recall their definition following mainly the work \cite{HW}.
 	
Let $d$-dimensional Euclidean space $E^d$ is divided into regularly spaced cells representing convex polyhedra.  Such a division is similar to splitting the axis at intervals of periodicity of the function $\cos{\phi}$.
We assume that there are many hyperplanes that divide any of these polyhedra on a finite number of
small congruent polyhedra, similar to the original one (in the classical one-dimensional case each hyperplane  degenerates
into the point that divides the interval of periodicity in two equal parts). Such polyhedra are called foldable.
The reflection of each polyhedron with respect to any hyperplane translates it into another polyhedron of splitting.
Such a cellular structure of folding polyhedra are defined by a corresponding group of reflections.

Let $L$ be a simple Lie algebra and $R$ be a reducible system of roots. A system of roots is a set of vectors in $d$-dimensional
Euclidean space $E^d$ supplied with a scalar product $(.,.)$ which is completely determined by a basis of simple roots
$\alpha_i,\,i=1,..,d$ and by a group of reflections of $R$ called Weyl the group $W(R)$.
Generating elements of the Weyl group $w_i,\,i=1,..,d$ act on any vector $x\in E^d$  according to the formula

\begin{equation}
\label{ic-3}
w_i\,x=x-\frac{2(x,\alpha_i)}{(\alpha_i,\alpha_i)}\alpha_i.
\end{equation}
Specifically, if $x = \alpha_i$ we obtain from (\ref{ic-3}) $w_i\,\alpha_i =-\alpha_i$.
To any root $\alpha$ from the system $R$ corresponds the coroot
$$
\alpha^{\vee}=\frac{2\alpha}{(\alpha,\alpha)}.
$$
For the basis of the simple coroots $\alpha^{\vee}_i,\,i=1,..,d$ one can define the dual basis of fundamental weights
$\lambda_i,\,i=1,..,d$
$$
(\lambda_i,\alpha^{\vee}_j)=\delta_{ij}
$$
(here and in what follows the dual space ${E^d}^*$ is identified with $E^d$).
The bases of roots and weights are related by the linear transformation
\begin{equation}\label{ic-4}
\alpha_i=\sum_jC_{ij}\lambda_j, \quad C_{ij}=\frac{2(\alpha_i,\alpha_j)}{(\alpha_j,\alpha_j)},
\end{equation}
where $C$ is the Cartan matrix of the Lie algebra $L$.

The set of weights $\lambda$ satisfying the inequality $(\lambda,\alpha_i)>0$, where $\alpha_i,\,i=1,..,d,$
are the simple roots, is called the dominant Weyl chamber $C_0$. The images of $C_0$ obtained
by action on it by all elements of the Weyl group are called the Weyl chambers or fundamental
domains. Thus the Weyl group acts on the set of Weyl chambers transitively. Weyl chambers play
the role of convex polyhedra in the partitioning of the space $E^d$ which was described above. A reflection
of a Weyl chamber with respect to any hyperplane reduces to composition of the
action of elements of the Weyl group and translation group. The semidirect product of a
Weyl group $W$ and a commutative translation group $T$ is called an affine Weyl group $\widetilde W$
of a root system $R$. The group $\widetilde W$ is just a symmetry group of the cell structure of the space $E^d$.

A function which can be considered as a possible candidate for a $d$ - dimensional generalization of
$\cos{n\phi}$, i.e., multivariable Chebyshev polynomial, must be invariant under the action of the
Weyl group and have the property of periodicity.  Following \cite{K1} - \cite{LU}, we define the
periodical function in $d$ variables (orbit function in terms of the work \cite{KP}) by the formula
\begin{equation}
\label{ic-5}
\Phi_{\bf n}({\boldsymbol{\phi}}) = 
\sum\limits_{w\in {\mbox{\footnotesize W}}}e^{2\pi\ri (\emph{w}\,{\bf n},{\boldsymbol{\phi}})},
\end{equation}
which is obviously $W$ - invariant $\Phi_{\tilde w\,\bf n}({\boldsymbol{\phi}})=\Phi_{\bf n}({\boldsymbol{\phi}}), \,\,\forall \tilde w\,\in W.$
Here $\bf n$ is expressed in the basis of fundamental weights $\{\lambda_i\}$ and ${\boldsymbol{\phi}}$ is expressed
in the dual basis of coroots $\{\alpha^\vee_i\}$
$$
{\bf n}=\sum_{i=1}^d\,n_i\lambda_i \quad n_i\in Z, \quad  {\boldsymbol{\phi}}=
\sum_{i=1}^d\,\phi_i\alpha_i^\vee \quad \phi_i\in [0,1).
$$
The function $\Phi_{\bf n}({\boldsymbol{\phi}})$ defined by the relation (\ref{ic-5}), with non-negative integer $n_i$
from the ${\bf n} = (n_1,...,n_d)$, up to a normalization gives the $d$-variable Chebyshev polynomials $\Phi_{n_1,...,n_d}$ of the first kind.

For the function $\Phi_{\bf n}({\boldsymbol{\phi}})$ we have the following multiplication rule
\begin{equation}
\label{ic-6}
\Phi_{\bf k}\Phi_{\bf s}=\sum\limits_{w\in W}\Phi_{w\,{\bf k}+{\bf s}}.
\end{equation}
This formula allows us to obtain the recurrent relations for the corresponding polynomials of several variables.

The function $\Phi_{\bf n}({\boldsymbol{\phi}})$ from (\ref{ic-5}) can be represented in the form
$$
\Phi_{\bf n}({\boldsymbol{\phi}}) =
\sum\limits_{k=1}^{|W|}e^{2\pi\ri \sum\limits_{i=1}^{d}f_{ki}(\phi)\,n_i},
$$
where $f_{ki}(\phi)$ are some linear functions of $(\phi_1,..,\phi_d)$ and $|W|$ is the order of the Weyl group
(that is the number of its elements). Let us introduce the following diagonal matrix
\begin{equation}
\label{ic-7}
M={\rm diag}\left(e^{2\pi\ri \sum\limits_{i=1}^{d}f_{1i}(\phi)\,n_i},...,e^{2\pi\ri \sum\limits_{i=1}^{d}f_{|W|i}(\phi)\,n_i}\right).
\end{equation}
In view of (\ref{ic-7}) we have
\begin{equation}\label{ic-8}
\Phi_{\bf n}({\boldsymbol{\phi}}) =
{\rm tr}\,M.
\end{equation}
The matrix $M$ can be rewritten as a product of diagonal matrices $M_k$ \begin{equation}\label{ic-9}
M=\prod\limits_{k=1}^dM_k^{n_k},
\end{equation}
where
\begin{equation}\label{ic-10}
M_k={\rm diag}\left(e^{2\pi\ri f_{1k}(\phi)},...,e^{2\pi\ri f_{|W|k}(\phi)}\right)=e^{2\pi\ri A_k},\quad
A_k={\rm diag}\left(f_{1k}(\phi),...,f_{|W|k}(\phi)\right).
\end{equation}
Now let us associate to each $M_k$ the following matrix
\begin{equation}\label{ic-11}
R_k=(I_{|W|}-p_kM_k)^{-1},
\end{equation}
where $I_{|W|}$ is the unit $|W|\times |W|$-matrix and $p_k$ is a real parameter. This matrix $R_k$ is different from
the resolvent matrix for $M_k$ in the location of the spectral parameter. Since all the matrices $R_k$ are diagonal, they
commute with each other. Obviously, the following relation is hold
\begin{equation}
\label{ic-12}
M_k^{n_k} = \frac1{n_k!}\left.\frac{\rd^{n_k}}{\rd p_k^{n_k}}\left(R_{p_k}
\right)\right|_{p_k = 0}.
\end{equation}
It follows from (\ref{ic-8}), (\ref{ic-9}) and (\ref{ic-12}) that
\begin{equation}
\label{ic-13}
\Phi_{{\bf n}}({\boldsymbol{\phi}})=\Phi_{{n_1,..,n_d}}({\boldsymbol{\phi}}) = 
{\rm tr}(M_1^{n_1}..M_n^{n_d}) = 
\frac1{n_1!..n_d!}\left.\frac{\partial ^{n_1+..+n_d}}{\partial p_1^{n_1}..\partial p_n^{n_d}}
\left({\rm tr}(R_{p_1}..R_{p_d})
\right)\right|_{p_1= .. =p_d = 0}.
\end{equation}
The following function of the parameters $\{p_i\}$
\begin{equation}
\label{ic-14}
F_{p_1,..,p_d}^I =
{\rm tr}(R_{p_1}..R_{p_d}) = \sum\limits_{n_1..n_d\ge 0}\Phi_{{n_1,..,n_d}}({\boldsymbol{\phi}})p_1^{n_1}..p_d^{n_d},
\end{equation}
can be considered as a (non-normalized) generating function of Chebyshev polynomials in $d$ variables,
although $F_{p_1,..,p_d}^I$ is still defined in terms of exponent $e^{2\pi\ri f(\phi)_{jk}}$.
The expression of ${\rm tr}(R_{p_1}..R_{p_d})$ in terms of the variables $x_i$ defined as
\begin{equation}
\label{ic-15}
x_i\,=\,\Phi_{{\bf e}_i}({\boldsymbol{\phi}}),\quad {\bf e}_i = (\overbrace{0,..,0}^{i-1},1,\overbrace{0,..,0}^{d-i}),
\end{equation}
is not too cumbersome. This will be illustrated in the following examples of the Lie algebras $A_2$, $C_2$ and $G_2$.
 	
Calculation of the generating function of Chebyshev polynomials in several variables of the second
kind requires a slight modification of the method described above.

Classical Chebyshev polynomials of the second kind $U_n(x)$ is defined by the relation
\begin{equation}\label{ic-16}
U_n(x) = \frac{\sin{(n+1)\phi}}{\sin{\phi}},\quad n\ge 0.
\end{equation}
They satisfy the same recurrence relations (\ref{ic-1}) as the polynomials of the first kind with
different initial conditions
\begin{equation}\label{ic-17}
U_0(x)=1, \quad U_1(x)= 2x.
\end{equation}
Generalization of polynomials (\ref{ic-16}) for the case of several variables
based on the Weyl character formula \cite{Hu}, \cite{FH}, from which it follows
\begin{equation}
\label{ic-18}
U_{\bf n}({\boldsymbol{\phi}})=\frac{\sum\limits_{w\in {\mbox{\footnotesize W}}}\det{w}\,\,e^{2\pi\ri (\emph{w}\,
({\bf n}+\boldsymbol{\rho}),{\boldsymbol{\phi}})}}{\sum\limits_{w\in {\mbox{\footnotesize W}}}\det{w}\,\,
e^{2\pi\ri (\emph{w}\,{\boldsymbol{\rho}},{\boldsymbol{\phi}})}},
\end{equation}
where $\det{w} = (-1)^{\ell (\emph{w})}$  and ${\ell}(\emph{w})$ is the minimal number of Weyl
group generating elements $w_i$ required for expressing $w$  as a product of  $w_i$.
$\boldsymbol{\rho}$ is the Weyl vector which is equal to the half-sum of the positive roots.
Using the formula (\ref{ic-4}) the Weyl vector can be expressed in the basis of fundamental weights.

The function
$$
\Phi_{\bf n}^{as} = \sum\limits_{w\in {\mbox{\footnotesize W}}}\det{w}\,\,e^{2\pi\ri (\emph{w}\,
({\bf n}+\boldsymbol{\rho}),{\boldsymbol{\phi}})},
$$
which stands in the numerator of (\ref{ic-18}), can be represented as the difference between the expressions
$\Phi_{\bf n}^{as+}$ and $\Phi_{\bf n}^{as-}$
$$
\Phi_{\bf n}^{as} = \Phi_{\bf n}^{as+} - \Phi_{\bf n}^{as-} = \sum\limits_{w\in {\mbox{\footnotesize W}},\,
\det{w}=1}\,e^{2\pi\ri (\emph{w}\,({\bf n}+\boldsymbol{\rho}),{\boldsymbol{\phi}})}-
\sum\limits_{w\in {\mbox{\footnotesize W}},\,\det{w}=-1}\,e^{2\pi\ri (\emph{w}\,({\bf n}+\boldsymbol{\rho}),
{\boldsymbol{\phi}})}.
$$
Further calculation repeats the scheme given above (\ref{ic-7}) - (\ref{ic-14}) for each of the functions
$\Phi_{\bf n}^{as\pm}$ separately. In the considered case we must use the relation
\begin{equation}
\label{ic-19}
\Phi_{\bf n}^{as} = \frac1{n_1!..n_d!}\left.\frac{\rd^{n}}{\rd^{n_1}p_1..\rd^{n_d}p_n}
\left({\rm tr}(R^+_{p_1}..R^+_{p_d}-R^-_{p_1}..R^-_{p_d})
\right)\right|_{p_1= .. =p_d = 0}
\end{equation}
instead of $\Phi_{\bf n}$ given by (\ref{ic-13}). In the formula (\ref{ic-19}) the matrices $R^{\pm}_{p_i}$ are used to represent functions $\Phi_{\bf n}^{as\pm}$. More details are presented in the examples below.

To obtain  Chebyshev polynomials of the second kind we must in accordance with the formula (\ref{ic-18})
divide the function $\Phi_{\bf n}^{as}$ by the singular element $\Phi_{\bf 0}^{as}$, where ${\bf 0} = (\overbrace{0,..,0}^{d})$.
Thus, the generating function of Chebyshev polynomials of the second kind has the form
\begin{equation}
\label{ic-20}
F_{p_1,..,p_d}^{II} = \frac{{\rm tr}(R^+_{p_1}..R^+_{p_d}-R^-_{p_1}..R^-_{p_d})}{\Phi_{\bf 0}}.
\end{equation}
As in the case of the first kind  polinomials, transition to the variables $x_i = U_{{\bf e}_i}({\boldsymbol{\phi}})$ in
$F_{p_1,..,p_d}^{II}$ does not require cumbersome calculations.

In the second part of this work we calculate the generating functions of the two variables Chebyshev polynomials
associated with the systems of simple roots for the Lie algebras $A_2$, $C_2$ and $G_2$.

\section{The case of Lie algebra $A_2$}
\subsection{The generating function for the polynomials of the first kind}

In this section we obtain the well known two variables Chebyshev polynomials associated with
the root system of the Lie algebra $A_2$ by the technique presented in the Section 1. These polynomials were introduced by Koornwinder in \cite{K1}, and their generating functions were found in the work \cite{DL} by another method.

The reduced root system $R$ of the Lie algebra $A_2$ includes two fundamental roots $\alpha _1,\,\alpha_2$,
positive root $\alpha _1+\alpha_2$ and the reflections of these roots. The action of the Weyl group $W(A_2)$ generating elements
$w_1,w_2$ on the fundamental roots $\alpha _1,\,\alpha_2$ has the form

\begin{equation}
\label{evd-1}
w_1\alpha_1=-\alpha_1,\quad w_1\alpha_2=\alpha_1+\alpha_2,\quad w_2\alpha_1=\alpha_1+\alpha_2,\quad w_2\alpha_2=-\alpha_2.
\end{equation}
Using the formula (\ref{ic-4}) and the Cartan matrix
$$
C_{A_2}=
\left(
\begin{array}{cc}
2 & -1 \\
-1 & 2 \\
\end{array}
\right),
$$
we rewrite the above relations for the fundamental weights
\begin{equation}
\label{evd-2}
w_1\lambda_1=\lambda_2-\lambda_1,\quad w_1\lambda_2=\lambda_2,\quad
w_2\lambda_1=\lambda_1,\quad w_2\lambda_2=\lambda_1-\lambda_2.
\end{equation}
The action of the others group elements on the fundamental weights is determined by their representation
by the generating ones
$$
w_3=w_1w_2,\quad w_4=w_2w_1,\quad w_5=w_1w_2w_1,\quad w_0=e.
$$
Moreover
$$
\det{w_1} = \det{w_2} =\det{w_5} = -1,
$$
and the determinants of the other elements of the Weyl group is equal to the unit. The Weyl vector has the form
$$
\rho = \alpha_1 + \alpha_2 = \lambda_1 + \lambda_2.
$$

Using these formulas and the following notations  ${\bf n}=m\lambda_1+n\lambda_2$, ${\boldsymbol{\phi}}=\phi\alpha^{\vee}_1+\psi\alpha^{\vee}_2$, we represent the $W(A_2)$-invariant function of two variables $\Phi_{m,n}$ (\ref{ic-5}) in the form

\begin{equation}
\label{ac-1}
\begin{split}
\Phi_{m,n} &= e^{2\pi\ri m\phi}e^{2\pi\ri n\psi}+e^{2\pi\ri m(\psi-\phi)}e^{2\pi\ri n\psi}+
e^{2\pi\ri m\phi}e^{2\pi\ri n(\phi-\psi)}+
e^{2\pi\ri m(\psi-\phi)}e^{-2\pi\ri n\phi}+{}\\
&+e^{-2\pi\ri m\psi}e^{2\pi\ri n(\phi-\psi)}+e^{-2\pi\ri m\psi}e^{-2\pi\ri n\phi}.
\end{split}
\end{equation}
In accordance with the formulas (\ref{ic-5}) - (\ref{ic-10}), we introduce the diagonal matrices
\begin{equation}
\label{ac-14}
\begin{split}
A_1&={\rm diag}(\phi,\psi-\phi,\phi,\psi-\phi,-\psi,-\psi),{}\\
A_2&={\rm diag}(\psi,\psi,\phi-\psi,-\phi,\phi-\psi,-\phi),
\end{split}
\end{equation}
and define the matrices $M_1$ and $M_2$
$$
M_k=e^{2\pi\ri\,A_k},\quad k=1,2.
$$
Then the function $\Phi_{m,n}$ can be written in the following form
$$
\Phi_{m,n}= {\rm tr}(M_1^mM_2^n).
$$
 	
Let us introduce the matrices
$$
R_1=(I_6-pM_1)^{-1},\, R_2=(I_6-qM_2)^{-1},
$$
where $I_6$ is the unit $6\times 6$-matrix and $p,\,q$ are the real parameters. In view of the relations
$$
\left.\frac1{m!}\frac{dR_1^m}{dp^m}\right|_{p = 0} = M_1^m,\quad \left.\frac1{n!}\frac{dR_2^n}{dq^n}\right|_{q = 0} =M_2^n,
$$
the function $\Phi_{m,n}$ can be expressed as
$$
\Phi_{m,n} = \frac1{m!n!}\left.\frac{{\partial}^{m+n}{\rm tr}(R_1R_2)}{{\partial}p^m{\partial}q^n}\right|_{p,q = 0}.
$$
The matrix $R_1R_2$ is diagonal and its trace has the form
\begin{multline*}
{\rm tr}(R_1\,R_2)=\frac1{(1-pe^{2\pi\ri\phi})(1-qe^{2\pi\ri\psi})}
                   +\frac1{(1-pe^{2\pi\ri(\psi-\phi)})(1-qe^{2\pi\ri\psi})}+\\
                   \frac1{(1-pe^{2\pi\ri\phi})(1-qe^{-2\pi\ri(\phi-\psi)})}+
                   +\frac1{(1-pe^{2\pi\ri(\psi-\phi)})(1-qe^{-2\pi\ri\phi})}+\\
                   \frac1{(1-pe^{-2\pi\ri\psi})(1-qe^{2\pi\ri(\phi-\psi)})}
                   +\frac1{(1-pe^{-2\pi\ri\psi})(1-qe^{2\pi\ri\phi})}.
\end{multline*}
Bringing the sum in the right hand side of this expression to a common denominator and
expressing the exponential coefficients before $p,q$ by the complex-conjugate variables $z,\, \bar z$
defined by the formula
\begin{equation}
\label{ac-2}
z=\frac1{2}\Phi_{1,0}=e^{2\pi\ri\phi}+e^{2\pi\ri (\psi-\phi)}+e^{-2\pi\ri\psi},\quad
\bar z=\frac1{2}\Phi_{0,1}=e^{2\pi\ri\psi}+e^{2\pi\ri (\phi-\psi)}+e^{-2\pi\ri\phi},
\end{equation}
we obtain the generating function of the Chebyshev polynomials of the first kind
associated with the Lie algebra $A_2$
\begin{equation}\label{ac-4}
F_{p,q}^I = \frac{\sum\limits_{i,j=0}^2K_{ij}p^iq^j}{(1-zp+\bar zp^2-p^3)(1-\bar zq+zq^2-q^3)}.
\end{equation}
The coefficients $K_{ij}$ in (\ref{ac-4}) are given by the formulas
\begin{eqnarray}
K_{00}&=&6,\nonumber\\
K_{10}&=&-4z,\nonumber\\
K_{20}&=&2\bar z,\nonumber\\
K_{01}&=&-4\bar z,\nonumber\\
K_{11}&=&3(z\bar z-1),\label{ac-5}\\
K_{21}&=&-2(\bar z^2-z),\nonumber\\
K_{02}&=&2z,\nonumber\\
K_{12}&=&-2(z^2-\bar z),\nonumber\\
K_{22}&=&z\bar z-3.\nonumber
\end{eqnarray}
The resulting function (\ref{ac-4}) with coefficients (\ref{ac-5}) coincides with the corresponding
generating function for Chebyshev polynomials of the first kind, obtained in \cite{DL} (theorem 3.3).

The polynomials calculated by (\ref{ac-4}) are different in normalization from the ones
introduced by Koornwinder in the works \cite{K1}. If we put
$$
T_{0,0} = \frac1{6}\,\Phi_{0,0};\quad T_{m,0} = \frac1{2}\,\Phi_{m,0};
\quad T_{0,n} = \frac1{2}\,\Phi_{0,n};\quad T_{m,n} = \Phi_{m,n},\quad mn\neq 0;
$$
we obtain exactly the Chebyshev polynomials of the first kind $T_{m,n}$ in the Koornwinder normalization.
Let us list a few polynomials, taking into account that $T_{m,n}=\overline{T}_{n,m}$,
\begin{eqnarray}
T_{0,0}&=&1,\nonumber\\
T_{1,0}&=&z,\nonumber\\
T_{2,0}&=&z^2-2\bar z,\nonumber\\
T_{1,1}&=&z\bar z-3,\nonumber\\
T_{3,0}&=&z^3-3z\bar z+3,\nonumber\\
T_{2,1}&=&z^2\bar z-2\bar z^2-z),\nonumber\\
T_{4,0}&=&z^4 -4z^2\bar z+2\bar z^2-2 +4z,\nonumber\\
T_{3,1}&=&z^3\bar z - 3z\bar z^2 - z^2 + 5\bar z,\nonumber\\
T_{2,2}&=&z^2\bar z^2 -2z^3 -2\bar z^3 +4z\bar z-3.\nonumber
\end{eqnarray}

The polynomials $\Phi_{m,n}$ satisfy recurrence relations of the form (\ref{ic-6})
\begin{eqnarray}
\Phi_{k+1,m}&=&\Phi_{1,0}\,\Phi_{k,m} -\Phi_{k,m-1}-\Phi_{k-1,m+1}\\
\Phi_{k,m+1}&=&\Phi_{0,1}\,\Phi_{k,m}-\Phi_{k-1,m}-\Phi_{k+1,m-1}.\label{ac-6}
\end{eqnarray}
These recurrence relation (\ref{ac-6}) can be reduced to the following "linear" form  by the simple
transformation
\begin{eqnarray}
\Phi_{k+1,m}=\Phi_{1,0}\,\Phi_{k,m}-\Phi_{0,1}\,\Phi_{k-1,m}+\Phi_{k-2,m},\label{ac-7}\\
\Phi_{k,m+1}=\Phi_{0,1}\,\Phi_{k,m}-\Phi_{1,0}\,\Phi_{k,m-1}+\Phi_{k,m-2}.\label{ac-8}
\end{eqnarray}
Using the relations of a type similar to (\ref{ac-7}), (\ref{ac-8}) one can to obtain  generating functions of Chebyshev polynomials by a scheme slightly changed from the present section. We will demonstrate it below in the section
dedicated to the polynomials associated with the root system of the Lie algebra $C_2$.

\subsection{The generating function for the polynomials of the second kind}

Now we calculate the generating function of the Chebyshev polynomials of the second kind associated with the root system of the Lie algebra $A_2$. Taking into account the form of the Weyl vector $\rho = \lambda_1+\lambda_2$ we rewrite the general formula (\ref{ic-18}) in the present case as

\begin{equation}\label{ac-9}
U_{\bf n}({\boldsymbol{\phi}})=\frac{\sum\limits_{w\in {\mbox{\footnotesize W}}}\det{w}\,e^{2\pi\ri (\emph{w}\,
({\bf n}+\boldsymbol{\rho}),{\boldsymbol{\phi}})}}{\sum\limits_{w\in {\mbox{\footnotesize W}}}\det{w}\,
e^{2\pi\ri (\emph{w}\,{\boldsymbol{\rho}},{\boldsymbol{\phi}})}} = \frac{\Phi_{\bf n+1}^{as}}{\Phi_{\bf 1}^{as}},
\end{equation}
where the numerator $\Phi_{\bf n+1}^{as}=\Phi_{m+1,n+1}^{as}$ and the denominator $\Phi_{1,1}^{as}$ of (\ref{ac-9}) are given by the following  expressions
\begin{equation}
\label{ac-10}
\begin{split}
\Phi_{m+1,n+1}^{as} &= (e^{2\pi\ri (m+1)\phi}e^{2\pi\ri (n+1)\psi}+
e^{2\pi\ri (m+1)(\psi-\phi)}e^{-2\pi\ri (n+1)\phi}
+e^{-2\pi\ri (m+1)\psi}e^{2\pi\ri (n+1)(\phi-\psi)})-{}\\
&-(e^{2\pi\ri (m+1)(\psi-\phi)}e^{2\pi\ri (n+1)\psi}+
e^{2\pi\ri (m+1)\phi}e^{2\pi\ri (n+1)(\phi-\psi)}+
e^{-2\pi\ri (m+1)\psi}e^{-2\pi\ri (n+1)\phi}),
\end{split}
\end{equation}
\begin{equation}\label{ac-11}
\Phi_{1,1}^{as} = (e^{2\pi\ri (\phi + \psi)}+e^{2\pi\ri (\psi-2\phi)}
+e^{2\pi\ri (\phi-2\psi)})-(e^{2\pi\ri(2\psi- \phi)}+
e^{2\pi\ri (2\phi-\psi)}+e^{-2\pi\ri (\phi + \psi)}),
\end{equation}
Let us introduce the following variables by
$$
U_{10} = \frac{\Phi_{2,1}^{as}}{\Phi_{1,1}^{as}},\quad U_{01} = \frac{\Phi_{1,2}^{as}}{\Phi_{1,1}^{as}}.
$$
Using the relations (\ref{ac-10}), (\ref{ac-11}), we obtain
$$
U_{10} =  z,\quad \quad U_{01} = \bar z,
$$
where $z$ and $\bar z$ are defined in (\ref{ac-2}).
 	
In accordance with the method described in the Introduction, we define the following four diagonal matrices
$$
M_{1+}={\rm diag}(e^{2\pi\ri \phi},e^{2\pi\ri(\psi-\phi)},e^{-2\pi\ri\psi}),\quad M_{2+}=
{\rm diag}(e^{2\pi\ri\psi},e^{-2\pi\ri \phi},e^{2\pi\ri(\phi-\psi)}),
$$
$$
M_{1-}={\rm diag}(e^{2\pi\ri(\psi-\phi)},e^{2\pi\ri \phi},e^{-2\pi\ri\psi}),\quad M_{2-}=
{\rm diag}(e^{2\pi\ri\psi},e^{2\pi\ri(\phi-\psi)},e^{-2\pi\ri \phi}).
$$
In terms of these matrices the function $\Phi_{m,n}^{as}$ can be rewritten in the form
$$
\Phi_{m,n}^{as} = {\rm tr}(M_{1+}^mM_{2+}^n-M_{1-}^mM_{2-}^n).
$$
Let us consider the matrices $R_{1\pm}=(I_6-pM_{1\pm})^{-1}$ and $R_{2\pm}=(I_6-qM_{2\pm})^{-1}$.
From the relation
$$\frac1{m!}\left.\frac{dR_{1\pm}^m}{dp^m}\right|_{p = 0} = M_{1\pm}^m,\quad
\frac1{n!}\left.\frac{dR_{2\pm}^n}{dp^n}\right|_{q = 0} =M_{2\pm}^n,$$
it follows that the function
$$
F^{II}_r(p,q) = \frac{{\rm tr}(R_{1+}R_{2+}-R_{1-}R_{2-})}{\Phi_{1,1}^{as}},
$$
is the generating function of the two variables Chebyshev polynomials of the second kind
$$
U_{m,n} = \frac1{(m+1)!(n+1)!}\left.\frac{{\partial}^{m+n+2}F^{II}_r(p,q)}{{\partial}p^{m+1}{\partial}q^{n+1}}\right|_{p = 0,q=0}.
$$
After a calculation of the traces
$$
{\rm tr}(R_{1+}R_{2+}) = \frac1{(1-pe^{iA})(1-qe^{iB})}+\frac1{(1-pe^{i(B-A)})(1-qe^{iA})}+
\frac1{(1-pe^{-iB)})(1-qe^{i(A-B)})},
$$
$$
{\rm tr}(R_{1-}R_{2-}) = \frac1{(1-pe^{i(B-A)})(1-qe^{iB})}+\frac1{(1-pe^{iA})(1-qe^{i(A-B)})}
+\frac1{(1-pe^{-iB)})(1-qe^{-iA)})}.
$$
and changing of the variables (\ref{ac-2}) we obtain
$$
 {\rm tr}(R_{1+}R_{2+}-R_{1-}R_{2-}) = \frac{pq(1-pq)\Phi_{1,1}^{as}}{(1-zp+\bar zp^2-p^3)(1-\bar zq+zq^2-q^3)}.
$$
Dividing this expression on the singular element $\Phi_{1,1}^{as}$ and on the product $pq$ (to ensure that polynomial subscripts
are in agreement with the order of differentiation) we obtain for the generating function
of the Chebyshev polynomials of the second kind the known formula \cite{DL}

\begin{equation}\label{ac-12}
F^{II}(p,q) = \frac{1-pq}{(1-zp+\bar zp^2-p^3)(1-\bar zq+zq^2-q^3)}.
\end{equation}
Using (\ref{ac-12}) we obtain by differentiation
\begin{equation}\label{ac-13}
U_{m,n} = \frac1{m!n!}\left.\frac{{\partial}^{m+n}F^{II}(p,q)}{{\partial}p^{m}{\partial}q^{n}}\right|_{p = 0,q=0}.
\end{equation}
exactly the polynomials given by Koornwinder \cite{K1} without additional normalization.
We list a few polynomials, taking into account that $U_{m,n}=\overline{U}_{n,m}$

\begin{eqnarray}
U_{0,0}&=&1,\nonumber\\
U_{1,0}&=&z,\nonumber\\
U_{2,0}&=&z^2-\bar z,\nonumber\\
U_{1,1}&=&z\bar z-1,\nonumber\\
U_{3,0}&=&z^3-2z\bar z+1,\nonumber\\
U_{2,1}&=&z^2\bar z-\bar z^2-z,\nonumber\\
U_{4,0}&=&z^4 -3z^2\bar z+\bar z^2+2z,\nonumber\\
U_{3,1}&=&z^3\bar z - 2z\bar z^2 - z^2 + 2\bar z,\nonumber\\
U_{2,2}&=&z^2\bar z^2 -z^3 -\bar z^3.\nonumber
\end{eqnarray}

\section{The case of Lie algebra $C_2$}
\subsection{The generating function for the polynomials of the first kind}

The root system of the Lie algebra $C_2$ has two fundamental roots $\alpha _1,\,\alpha_2$ and includes the positive roots
$\alpha _1+\alpha_2,\,2\alpha _1+\alpha_2$ together with their reflections. Using the formula (\ref{ic-3})
and the Cartan matrix of the algebra $C_2$
$$
C_{C_2}=\left(\begin{array}{cc}
2 & -1 \\
-2 & 2 \\
\end{array}\right),
$$
we obtain the action of the Weyl group $W(C_2)$ generating elements $w_1,w_2$ on the fundamental roots
$$
w_1\alpha_1=-\alpha_1,\quad w_1\alpha_2=2\alpha_1+\alpha_2,\quad w_2\alpha_1=\alpha_1+\alpha_2,\quad w_2\alpha_2=-\alpha_2.
$$
Next we rewrite these relations for the fundamental weights by means of the formula (\ref{ic-4})
$$
w_1\lambda_1=\lambda_2-\lambda_1,\quad w_1\lambda_2=\lambda_2,\quad w_2\lambda_1=\lambda_1,
\quad w_2\lambda_2=2\lambda_1-\lambda_2.
$$
The action of the others group elements on the fundamental weights is determined by their representation
by the generating elements
\begin{equation}
\label{cc-1}
w_3=w_1w_2,\quad w_4=w_2w_3,\quad w_5=w_1w_2w_1,\quad w_6=w_2w_1w_2,\quad w_7=(w_1w_2)^2,\quad e=w_0.
\end{equation}
The determinants of the Weyl group elements are given by
$$
{\rm det}w_1 = {\rm det}w_2 = {\rm det}w_5 ={\rm det}w_6 = -1,
$$
and the others are equal to the unity. The Weyl vector has the form
\begin{equation}\label{cc-2}
\rho = 2\alpha_1 + \frac{3}{2}\alpha_2 = \lambda_1+\lambda_2.
\end{equation}

Repeating the calculations from the Section 2.1 and using the notations which were introduced there
we find $W(C_2)$ - invariant function (orbit function) in two variables
\begin{equation}\label{cc-3}
\begin{split}
\Phi_{m,n} &= e^{2\pi\ri(m\phi+n\psi)}+e^{2\pi\ri(m(\psi-\phi)+n\psi))}+
e^{2\pi\ri(m\phi+n(2\phi-\psi))}+e^{2\pi\ri(m(\psi-\phi)+n(-2\phi+\psi))}+  {}\\
&+ e^{2\pi\ri(m(\phi-\psi)+n(2\phi-\psi))}+e^{2\pi\ri(-m\phi+n(-2\phi+\psi))}+
e^{2\pi\ri(m(\phi-\psi)-n\psi)}+e^{2\pi\ri(-m\phi-n\psi)}.
\end{split}
\end{equation}
Note that the function $\Phi_{m,n}$ (\ref{cc-3}) is real. The matrices $A_i$ (\ref{ac-14}) in the considered case have the form
$$
A_1={\rm diag}(\phi,\psi-\phi,\phi,\psi-\phi,\phi-\psi,-\phi,\phi-\psi,-\phi)
$$
and
$$
A_2={\rm diag}(\psi,\psi,2\phi-\psi,-2\phi+\psi,2\phi-\psi,-2\phi+\psi,-\psi,-\psi).
$$
Then the function $\Phi_{m,n}$ can be expressed by the matrices $M_k=e^{2\pi\ri\,A_k},\quad k=1,2$
in the trace form
$$
\Phi_{m,n}= {\rm tr}(M_1^mM_2^n).
$$
Following \cite{KP}, we define the Chebyshev polynomials of the first kind for algebra $C_2$
by the relations
\begin{equation}\label{cc-4}
T_{0,0}=\frac1{8}\Phi_{0,0},\quad T_{m,0}=\frac1{2}\Phi_{m,0},\quad
T_{0,n}=\frac1{2}\Phi_{0,n},\quad T_{m,n} = \Phi_{m,n}\quad m\cdot n\neq 0.
\end{equation}
With this definition $T_{0,0}=1$ and the new real variables $x,y$ are given by the formulas
\begin{eqnarray}
x&=&T_{1,0}=e^{2\pi\ri\phi}+e^{-2\pi\ri\phi}+ e^{2\pi\ri(\phi-\psi)}+e^{-2\pi\ri(\phi-\psi)}, \label{cc-5}\\
y&=&T_{0,1}=e^{2\pi\ri\psi}+e^{-2\pi\ri\psi}+e^{2\pi\ri(2\phi-\psi)}+ e^{-2\pi\ri(2\phi-\psi)}. \label{cc-6}
\end{eqnarray}

Turning to the calculation of the generating function we introduce the matrices
$$
R_p = (I_8-pM_1)^{-1},\quad R_q = (I_8-qM_2)^{-1},
$$
which are diagonal with the elements given by
$$
\left(1-p\exp{(2\pi\ri{(A_1)}_{kk})}\right)^{-1},\quad \left(1-q\exp{(2\pi\ri{(A_2)}_{kk})}\right)^{-1},
$$
where $(A_i)_{kk}$ are the diagonal elements of the matrices $A_i,\,i=1,2,$ and $I_8$ is the unit $8\times 8$ matrix.
Substituting $R_p,R_q$ in the equation (\ref{ic-13}) and expressing the coefficients of $p,q$ as
functions of $x$ (\ref{cc-5}) and $y$ (\ref{cc-6}), we obtain from (\ref{ic-14})
\begin{equation}
\label{cc-7}
F_{p,q}^I = \frac{\sum\limits_{i,j=0}^3K_{ij}p^iq^j}{(1-xp+(2+y)p^2-xp^3+p^4)(1-yq+(x^2-2y-2)q^2-yq^3+q^4)},
\end{equation}
	
where $K_{ij}$ are given by the formulas
\begin{eqnarray}
K_{00}&=&8,\nonumber\\
K_{10}&=&-6x,\nonumber\\
K_{20}&=&4y+8,\nonumber\\
K_{30}&=&-2x,\nonumber\\
K_{01}&=&-6y,\nonumber\\
K_{11}&=&5xy-2x,\nonumber\\
K_{21}&=&-4y^2+2x^2-10y,\nonumber\\
K_{31}&=&2xy-2x,\label{cc-8}\\
K_{02}&=&4x^2-8y-8,\nonumber\\
K_{12}&=&-4x^3+9xy+10x,\nonumber\\
K_{22}&=&3x^2y-6y^2+4x^2-20y-8,\nonumber\\
K_{32}&=&-2x^3+5xy+6x,\nonumber\\
K_{03}&=&-2y,\nonumber\\
K_{13}&=&2xy-2x,\nonumber\\
K_{23}&=&-2y^2+2x^2-6y,\nonumber\\
K_{33}&=&xy-2x.\nonumber
\end{eqnarray}
A few  polynomials calculated  using (\ref{cc-7}) with the normalization (\ref{cc-4}) are listed below
\begin{eqnarray}
T_{0,0}&=&1,\nonumber\\
T_{1,0}&=&x,\nonumber\\
T_{2,0}&=&x^2-2y-4,\nonumber\\
T_{3,0}&=&x^3-3xy-3x,\nonumber\\
T_{0,1}&=&y,\nonumber\\
T_{1,1}&=&xy-2x,\nonumber\\
T_{2,1}&=&x^2y-2y^2-6y,\nonumber\\
T_{3,1}&=&x^3y-3xy^2-4xy+2x,\label{cc-9}\\
T_{0,2}&=&y^2-2x^2+4y+4,\nonumber\\
T_{1,2}&=&xy^2-2x^3+3xy+6x,\nonumber\\
T_{2,2}&=&x^2y^2-2x^4-2y^3-12y^2+8x^2y+10x^2-20y-8,\nonumber\\
T_{3,2}&=&x^3y^2-2x^5-3xy^3+10x^3y-15y^2x+10x^3-25xy-10x,\nonumber\\
T_{0,3}&=&y^3-3x^2y+6y^2+9y,\nonumber\\
T_{1,3}&=&xy^3-3x^3y+5y^2x+2x^3+6xy-6x,\nonumber\\
T_{2,3}&=&x^2y^3-2y^4-3x^4y-16y^3+12x^2y^2+20x^2y-40y^2-30y,\nonumber\\
T_{3,3}&=&x^3y^3-3x^5y-3xy^4+15x^3y^2-21xy^3+18x^3y-45y^2x-2x^3-21xy+6x.\nonumber
\end{eqnarray}

The recurrence relations for the polynomials under consideration can be obtained by
the multiplication rules (\ref{ic-6}) putting ${\bf s}=(m,n)$ and ${\bf k}=(1,0)$ in the first relation
and ${\bf k}=(0,1)$ in the second one. As a result we finally obtain
\begin{eqnarray}
\Phi_{1,0}\Phi_{m,n}&=&\Phi_{m+1,n}+\Phi_{m-1,n}+\Phi_{m+1,n-1}+\Phi_{m-1,n+1}, \label{cc-10} \\
\Phi_{0,1}\Phi_{m,n}&=&\Phi_{m,n+1}+\Phi_{m,n-1}+\Phi_{m+2,n-1}+\Phi_{m-2,n+1}. \label{cc-11}
\end{eqnarray}

For our purposes the relations  (\ref{cc-10}), (\ref{cc-11}) must be reduced to the "linear"$\,$ form
(see (\ref{ac-7}), (\ref{ac-8})). Let us show the simple way to do this reduction.

The determinant of the matrix $pI_8-M_1$, gives us the characteristic equation, which satisfies the matrix $M_1$.
Since all the eigenvalues of this matrix have multiplicity 2, the minimal polynomial of $M_1$
(after transition to the variables $x,y$) takes the form
$$
P_1=1-xp+(2+y)p^2-xp^3+p^4.
$$
Therefore, the matrix $M_1$ satisfy the equation
$$
M_1^4-xM_1^3+(2+y)M_1^2-xM_1+I_8=0.
$$
Multiplying this equation from the right by $M_1^{m-4}M_2^n$ and taking the trace we obtain
\begin{equation}\label{cc-12}
\Phi_{m,n} = x\Phi_{m-1,n}-(2+y)\Phi_{m-2,n}+x\Phi_{m-3,n}-\Phi_{m-4,n}.
\end{equation}
Acting in the same way with the matrix $M_2$, we obtain the characteristic polynomial
$$
P_2=1-yq+(x^2-2y-2)q^2-yq^3+q^4
$$
and the second recurrent relation
\begin{equation}\label{cc-13}
\Phi_{m,n} =y\Phi_{m,n-1}-(x^2-2y-2)\Phi_{m,n-2}+y\Phi_{m,n-3}-\Phi_{m,n-4}.
\end{equation}
The linear relations (\ref{cc-12}) and (\ref{cc-13}) can be tested independently by
direct substitution of (\ref{cc-10}) and (\ref{cc-11}).

The relations (\ref{cc-12}) and (\ref{cc-13}) allow us to find the generating function (\ref{cc-7}) somewhat differently than it was done above. Let us rewrite them in the following form
\begin{equation}
\label{cc-14}
\Phi_{m,n} =\left(
           \begin{array}{cccc}
             \Phi_{m-1,n}\,; & -\Phi_{m-2,n}\,; & \Phi_{m-3,n}\,; & -\Phi_{m-4,n}
           \end{array}
         \right)
\left(
  \begin{array}{c}
    x \\
    y+2 \\
    x \\
    1 \\
  \end{array}
\right),
\end{equation}
\begin{equation}
\label{cc-15}
\Phi_{m,n} =\left(
           \begin{array}{cccc}
             \Phi_{m,n-1}\,; & -\Phi_{m,n-2}\,; & \Phi_{m,n-3}\,; & -\Phi_{m,n-4}
           \end{array}
         \right)
\left(
  \begin{array}{c}
    y \\
    x^2-2y-2\\
    y \\
    1 \\
  \end{array}
\right).
\end{equation}
By means of the iteration of the relations (\ref{cc-14}) and (\ref{cc-15}) it is not difficult to transform them to the form

\begin{equation}
\label{cc-16}
\Phi_{m,n} =\left(
           \begin{array}{cccc}
             \Phi_{m-1-k,n}\,; & -\Phi_{m-2-k,n}\,; & \Phi_{m-3-k,n}\,; & -\Phi_{m-4-k,n} \\
           \end{array}
         \right)
M_x^k
\left(
  \begin{array}{c}
    x \\
    y+2 \\
    x \\
    1 \\
  \end{array}
\right),
\end{equation}
\begin{equation}
\label{cc-17}
\Phi_{m,n} =\left(
           \begin{array}{cccc}
             \Phi_{m,n-1-k}\,; & -\Phi_{m,n-2-k}\,; & \Phi_{m,n-3-k}\,; & -\Phi_{m,n-4-k} \\
           \end{array}
         \right)
M_y^k
\left(
  \begin{array}{c}
    y \\
    x^2-2y-2\\
    y \\
    1 \\
  \end{array}
\right),
\end{equation}
where
$$
M_x =
\left(
  \begin{array}{cccc}
    x & -1 & 0 & 0 \\
    y+2 & 0 & -1 & 0 \\
    y & 0 & 0 & -1 \\
    1 & 0 & 0 & 0 \\
  \end{array}
\right),\quad
M_y=\left(
  \begin{array}{cccc}
    y & -1 & 0 & 0 \\
    x^2-2y-2 & 0 & -1 & 0 \\
    y & 0 & 0 & -1 \\
    1 & 0 & 0 & 0 \\
  \end{array}
\right).
$$
Substitute $k=m-4$  in (\ref{cc-16}) and $k=n-4$ in (\ref{cc-17}). We find
$$
\Phi_{m,n} =\left(
           \begin{array}{cccc}
             \Phi_{3,n}\,; & -\Phi_{2,n}\,; & \Phi_{1,n}\,; & -\Phi_{0,n} \\
           \end{array}
         \right)
M_x^{m-4}
\left(
  \begin{array}{c}
    x \\
    y+2 \\
    x \\
    1 \\
  \end{array}
\right),
$$
$$
\Phi_{m,n} =\left(
           \begin{array}{cccc}
             \Phi_{m,3}\,; & -\Phi_{m,2}\,; & \Phi_{m,1}\,; & -\Phi_{m,0} \\
           \end{array}
         \right)
M_y^{n-4}
\left(
  \begin{array}{c}
    y \\
    x^2-2y-2\\
    y \\
    1 \\
  \end{array}
\right).$$
Because of the matrices $M_x,\,M_y$ are inverted (${\rm det}(M_x)={\rm det}(M_y)=1$) we have
$$
M_x^{-4}
\left(
  \begin{array}{c}
    x \\
    y+2 \\
    x \\
    1 \\
  \end{array}
\right)=
M_y^{-4}
\left(
  \begin{array}{c}
    y \\
    x^2-2y-2\\
    y \\
    1 \\
  \end{array}
\right)=
\left(
  \begin{array}{c}
    0 \\
    0\\
    0 \\
    -1 \\
  \end{array}
\right).
$$
Therefore, the relations (\ref{cc-16}), (\ref{cc-17}) can be converted to the final form
\begin{equation}\label{cc-18}
\Phi_{m,n} =\left(
           \begin{array}{cccc}
             \Phi_{3,n}\,; & -\Phi_{2,n}\,; & \Phi_{1,n}\,; & -\Phi_{0,n}
           \end{array}
         \right)
M_x^{m}
\left(
  \begin{array}{c}
    0 \\
    0 \\
    0 \\
    -1 \\
  \end{array}
\right),
\end{equation}
\begin{equation}\label{cc-19}
\Phi_{m,n} =\left(
           \begin{array}{cccc}
             \Phi_{m,3}\,; & -\Phi_{m,2}\,; & \Phi_{m,1}\,; & -\Phi_{m,0} \\
           \end{array}
         \right)
M_y^{n}
\left(
  \begin{array}{c}
    0 \\
    0 \\
    0 \\
    -1 \\
  \end{array}
\right).
\end{equation}
The obtained relations (\ref{cc-18}) and (\ref{cc-19}) allows us to represent any Chebyshev polynomial $\Phi_{m,n}$ with subscripts $m,n > 3$ by the polynomials with $0\leq m,n \leq 3$. Indeed, using the relation (\ref{cc-18}) with the polynomials $\Phi_{m,n},\,\,0\leq m,n \leq 3$, one can to find the polynomials $\Phi_{m,0},\Phi_{m,1},\Phi_{m,2},$ $\Phi_{m,3}$ with arbitrary subscript $m$. Then, using the relation (\ref{cc-19}) with the calculated polynomials, one can to find a polynomial $\Phi_{m,n}$ with an arbitrary  subscript $n$.

The considered modification of the method gives the advantage of the initial polynomials arbitrariness, that in turn allows us to construct different series of polynomials which satisfy linear recurrence relations (\ref{cc-12}) and (\ref{cc-13}), but generally speaking not the original relations (\ref{cc-10}) and (\ref{cc-11}). In other words, the arbitrariness in the assignment of initial polynomials is limited.

\subsection{The generating function for polynomials of second kind}

In the present Section we mainly follow the Introduction and Section 2.2. Because of the Weyl vector for the algebra $C_2$ is equal to $\rho = \lambda_1+\lambda_2$, the generating function for the Chebyshev polynomials of the second kind has the form (\ref{ac-9})
$$
U_{\bf n}({\boldsymbol{\phi}})=\frac{\sum\limits_{w\in {\mbox{\footnotesize W}}}\det{w}\,
e^{2\pi\ri (\emph{w}\,({\bf n}+\boldsymbol{\rho}),{\boldsymbol{\phi}})}}
{\sum\limits_{w\in {\mbox{\footnotesize W}}}\det{w}\,e^{2\pi\ri (\emph{w}\,
{\boldsymbol{\rho}},{\boldsymbol{\phi}})}} =
\frac{\Phi_{\bf n+1}^{as}}{\Phi_{\bf 1}^{as}}.
$$
Using the notations of Section 2.2 and the presentation of the Weyl group $W(C_2)$ elements by the generating ones of Section 3.1, we obtain for $\Phi_{m,n}^{as}$ and the singular element  $\Phi_{1,1}^{as}$ respectively

\begin{equation}
\label{cc-20}
\begin{split}
\Phi_{m,n}^{as} &= \left(e^{2\pi\ri(m\phi+n\psi)}
+e^{2\pi\ri(m(\psi-\phi)+n(-2\phi+\psi))}
+e^{2\pi\ri(-m\phi-n\psi)}
+ e^{2\pi\ri(m(\phi-\psi)+n(2\phi-\psi))}\right)
-  {}\\
&-\left(e^{2\pi\ri(m(\psi-\phi)+n\psi))}+
e^{2\pi\ri(m\phi+n(2\phi-\psi))}
+e^{2\pi\ri(-m\phi+n(-2\phi+\psi))}
+e^{2\pi\ri(m(\phi-\psi)-n\psi)}\right),
\end{split}
\end{equation}
\begin{equation}
\label{cc-21}
\begin{split}
\Phi_{1,1}^{as} &= (e^{2\pi\ri (\phi + \psi)}+e^{2\pi\ri (2\psi-3\phi)}
+e^{2\pi\ri (-\phi-\psi)}+e^{2\pi\ri (-2\psi+3\phi)})-{}\\
&-(e^{2\pi\ri(2\psi- \phi)}+e^{2\pi\ri (3\phi-\psi)}+e^{-2\pi\ri (-3\phi + \psi)}+e^{2\pi\ri (\phi-2\psi)}).
\end{split}
\end{equation}
We not introduce at this stage new variables by the formulas
$$
U_{10} = \frac{\Phi_{2,1}^{as}}{\Phi_{1,1}^{as}},\quad U_{01} = \frac{\Phi_{1,2}^{as}}{\Phi_{1,1}^{as}},
$$
because it will be more suitable to do later. Using the formulas (\ref{cc-20}), (\ref{cc-21}) we define the following four diagonal matrices
$$M_{1+}=
{\rm diag}(e^{2\pi\ri\phi},e^{2\pi\ri(\psi-\phi)},e^{-2\pi\ri\phi},e^{2\pi\ri(\phi-\psi)}),
\quad
M_{2+}=
{\rm
diag}(e^{2\pi\ri\psi},e^{2\pi\ri(-2\phi+\psi},e^{-2\pi\ri\psi},e^{2\pi\ri(2\phi-\psi}),
$$
$$
M_{1-}={\rm diag}
(e^{2\pi\ri(\psi-\phi)},e^{2\pi\ri \phi},e^{-2\pi\ri\phi},e^{2\pi\ri(\phi-\psi)}),
\quad
M_{2-}={\rm diag}
(e^{2\pi\ri\psi},e^{2\pi\ri(2\phi-\psi)},e^{2\pi\ri (-2\phi+\psi)},e^{-2\pi\ri\psi}).
$$
In terms of these matrices the function $\Phi_{m,n}^{as}$ has the form
$$
\Phi_{m,n}^{as} = {\rm tr}(M_{1+}^mM_{2+}^n-M_{1-}^mM_{2-}^n).
$$
Introducing the matrices
$R_{1\pm}=(I_4-pM_{1\pm})^{-1}$, $R_{2\pm}=(I_4-qM_{2\pm})^{-1}$,
one can to say that the function
$$
F^{II}_r(p,q) = \frac{{\rm tr}(R_{1+}R_{2+}-R_{1-}R_{2-})}{\Phi_{1,1}^{as}},
$$
is the generating function of the two variables Chebyshev polynomials of the second kind
$$
U_{m,n} = \frac1{(m+1)!(n+1)!}
\left.\frac{{\partial}^{m+n+2}F^{II}_r(p,q)}
{{\partial}p^{m+1}{\partial}q^{n+1}}\right|_{p = 0,q=0}.
$$
As a result of the simple calculations we obtain
$${\rm tr}(R_{1+}R_{2+}-R_{1-}R_{2-}) = \frac{pq(1+q+p^2q+p^2q^2-\tilde xpq)\Phi_{1,1}^{as}}{(1-\tilde xp+
(\tilde y+2)p^2-\tilde xp^3+p^4)(1-\tilde yq+(\tilde x^2-2\tilde y-2)q^2-\tilde yq^3+q^4)},$$
where was introduced the following intermediate notation
$$
\tilde x = {\rm tr}(M_{1+})={\rm tr}(M_{1-})=
e^{2\pi\ri\phi}+e^{2\pi\ri(\psi-\phi)}+e^{-2\pi\ri\phi}
+e^{2\pi\ri(\phi-\psi)},
$$
$$
\tilde y = {\rm tr}(M_{2+})={\rm tr}(M_{2-})=e^{2\pi\ri\psi}+e^{2\pi\ri(-2\phi+\psi}+e^{-2\pi\ri\psi}+
e^{2\pi\ri(2\phi-\psi}.
$$
Now as in the $A_2$ case we divide the obtained expression by the singular element $\Phi_{11}^{as}$ and
by the product $pq$. The calculation of derivatives gives us
$$
U_{10} = \left.\frac{dF^{II}_r(p,q)}{dp}\right|_{p = 0,q=0}= \tilde x ,\quad
U_{01} = \left.\frac{dF^{II}_r(p,q)}{dq}\right|_{p = 0,q=0}= \tilde y +1.
$$
Just in this place it is convenient to introduce the new variables
$$
x=\tilde x,\quad y = \tilde y + 1.
$$
In the terms of these variables $F^{II}(p,q)$ has the form
\begin{equation}\label{cc-22}
F^{II}(p,q) = \frac{1+q+p^2q+p^2q^2-xpq}{(1-xp+(y+1)p^2-xp^3+p^4)(1-(y-1)q+(x^2-2y)q^2-(y-1)q^3+q^4)}.
\end{equation}
Using the formula
$$
U_{m,n} = \frac1{m!n!}\left.\frac{{\partial}^{m+n}F^{II}(p,q)}{{\partial}p^{m}{\partial}q^{n}}\right|_{p = 0,q=0}
$$
we calculate a few  polynomials of the second kind
\begin{eqnarray}
U_{0,0}&=&1,\nonumber\\
U_{1,0}&=&x,\nonumber\\
U_{0,1}&=&y,\nonumber\\
U_{2,0}&=&x^2-y-1,\nonumber\\
U_{1,1}&=&xy-x,\nonumber\\
U_{0,2}&=&-x^2+y^2+y,\nonumber\\
U_{3,0}&=&x^3-2xy-x,\nonumber\\
U_{2,1}&=&x^2y-x^2-y^2-y+1,\nonumber\\
U_{1,2}&=&-x^3+xy^2+x,\nonumber\\
U_{0,3}&=&y^3-2x^2y+x^2+2y^2-1.\nonumber
\end{eqnarray}
These polynomials coincide with the ones which were found in \cite{LU} from the recurrence relations if interchange the polynomial subscripts and perform the following substitution $x\rightarrow X_2,\,\,y \rightarrow X_1.$

\section{The case of the Lie algebra $G_2$}

In this Section we calculate the generating function of the Chebyshev polynomials of the first kind associated with the exclusive Lie algebra $G_2$.

The set of the positive roots of the algebra $G_2$ root system includes two fundamental roots $\alpha _1,\,\alpha_2$ together with $\alpha _1+\alpha_2,\,2\alpha _1+\alpha_2,\,3\alpha _1+\alpha_2,\,3\alpha _1+2\alpha_2$.
Using the relation (\ref{ic-3}) and the Cartan matrix
$$C_{G_2}=\left(
\begin{array}{cc}
2 & -1 \\
-3 & 2 \\
\end{array}
\right),$$
we obtain the action of the Weyl group $W(G_2)$ generating elements $w_1,w_2$ on the fundamental roots
$$
w_1\alpha_1=-\alpha_1,\quad w_1\alpha_2=3\alpha_1+\alpha_2,\quad w_2\alpha_1=\alpha_1+\alpha_2,\quad w_2\alpha_2=-\alpha_2.
$$
Taking into account (\ref{ic-4}) we obtain for the fundamental weights
$$
w_1\lambda_1=\lambda_2-\lambda_1,\quad w_1\lambda_2=\lambda_2,\quad
w_2\lambda_1=\lambda_1,\quad w_2\lambda_2=2\lambda_1-\lambda_2.
$$
The simple calculation results in the following W-invariant functions corresponding to the group $W(G_2)$
\begin{equation}\label{g-01}
\begin{split}
&\Phi_{m,n}(x) = e^{2\pi\ri ((2m + n)\phi + (m + 2/3n)\psi)} +
               e^{-2\pi\ri ((2m + n)\phi + (m + 2/3n)\psi)} +
               e^{2\pi\ri ((m + n)\phi + (m + 2/3n)\psi)} +  {}\\
             & e^{-2\pi\ri ((m + n)\phi + (m + 2/3n)\psi)} +
               e^{2\pi\ri ((2m + n)\phi + (m + 1/3n)\psi)} +
               e^{-2\pi\ri ((2m + n)\phi + (m + 1/3n)\psi)} +
               e^{2\pi\ri ((m + n)\phi + 1/3n\psi)} + {}\\
              &e^{-2\pi\ri ((m + n)\phi + 1/3n\psi)} +
               e^{2\pi\ri (m\phi + (m + 1/3n)\psi)} +
               e^{-2\pi\ri (m\phi + (m + 1/3n)\psi)} +
               e^{2\pi\ri (m\phi - 1/3n\psi)} +
               e^{-2\pi\ri (m\phi - 1/3n\psi)}.
\end{split}
\end{equation}
This formula is consistent with the orbit function which was obtained in \cite{KP}.
 	
Let us introduce the diagonal matrices
$$
A_1={\rm diag}(2\phi+\psi,-2\phi-\psi,\phi+\psi,-\phi-\psi,2\phi+\psi,-2\phi-\psi,\phi,-\phi,\phi+\psi,-\phi-\psi,\phi,-\phi),
$$
$$
A_2={\rm diag}(\phi+2/3\psi,-\phi-2/3\psi,\phi+2/3\psi,-\phi-2/3\psi,\phi+1/3\psi,-\phi-1/3\psi,
\phi+1/3\psi,-\phi-1/3\psi,
$$
$$
+1/3\psi,-1/3\psi,1/3\psi,-1/3\psi).
$$
Then, as in the case of the algebra $C_2$, we define
$$
M_k=e^{2\pi\ri\,A_k},\quad k=1,2,\quad \Phi_{m,n}= {\rm tr}(M_1^mM_2^n).
$$
We introduce by definition the new real variables

\begin{equation}\label{g-02}
x = {\rm tr}(M_1)/2 = e^{2\pi\ri (2\phi + \psi)} +e^{-2\pi\ri (2\phi + \psi)} +e^{2\pi\ri (\phi + \psi)} +
e^{-2\pi\ri (\phi + \psi)} +e^{2\pi\ri\phi} +e^{-2\pi\ri\phi},
\end{equation}
and
\begin{equation*}
y = {\rm tr}(M_2)/2 = e^{2\pi\ri (\phi + 2/3\psi)} +e^{-2\pi\ri (\phi + 2/3\psi)} + e^{2\pi\ri (\phi + 1/3\psi)} +
e^{-2\pi\ri (\phi + 1/3\psi)} + e^{2/3\pi\ri\psi} +e^{-2/3\pi\ri\psi}.
\end{equation*}
Since all eigenvalues of the matrices $M_i,\,i=1,2$ in the considered case have multiplicity 2,
then the degree of its minimal polynomial is equal to 6
\begin{equation}\label{g-03}
\begin{split}
P_1&=1-xp+(y^3-3xy-5x-9y-9)p^2-(-2y^3+x^2+6xy+12x+18y+20)p^3+ {}\\
&+(y^3-3xy-5x-9y-9)p^4-xp^5+p^6,
\end{split}
\end{equation}
\begin{equation}\label{g-04}
P_2=1-yq+(y+x+3)q^2-(y^2-2x-4)q^3+(y+x+3)q^4-yq^5+q^6.
\end{equation}
Using the method described in Section 3 we obtain a linear recurrent relations
\begin{multline*}
\Phi_{m,n}-x\Phi_{m-1,n}+(y^3-3xy-5x-9y-9)\Phi_{m-2,n}-\\
(-2y^3+x^2+6xy+12x+18y+20)\Phi_{m-3,n}+ \\
(y^3-3xy-5x-9y-9)\Phi_{m-4,n}-x\Phi_{m-5,n}+\Phi_{m-6,n}=0,
\end{multline*}
\begin{equation*}
\Phi_{m,n}-y\Phi_{m,n-1}+(y+x+3)\Phi_{m,n-2}-(y^2-2x-4)\Phi_{m,n-3}+
(y+x+3)\Phi_{m,n-4}-y\Phi_{m,n-5}+\Phi_{m,n-6}=0.
\end{equation*}	
Acting by the above technique, we obtain the desired generating function in the form
\begin{equation}
\label{dks-5}
F_{p,q} = (P_1P_2)^{-1}\left(\sum\limits_{i,j=0}^5K_{ij}p^iq^j\right),
\end{equation}
where the polynomials $P_1,\,P_2$ are given by the relations (\ref{g-03}), (\ref{g-04}),
and the coefficients $K_{ij}$ have the form
\begin{eqnarray*}
K_{00} &=& 12,\nonumber\\
K_{10} &=&-10x,\nonumber\\
K_{20} &=&= 8y^3-24xy-40x-72y-72,\nonumber\\
K_{30} &=& -6x^2+12y^3-36xy-72x-108y-120,\nonumber\\
K_{40} &=& 48y^3-128x8y-208x-368y-36,\nonumber\\
K_{50} &=& -2x,\nonumber\\
K_{01} &=& -10y,\nonumber\\
K_{11}&=& 9xy+2y-2y^2+4x+12,\nonumber\\
K_{21} &=& 40xy-8y^4+25xy^2-2x^2+72y^2-6x+74y,\nonumber\\
K_{31} &=& 6x^2y-12y^4+35xy^2+2x^2+72xy+108y^2+6x+118y,\nonumber\\
K_{41} &=& -4y^4+12xy^2+21xy+38y^2-4x+34y-12,\nonumber\\
K_{51} &=& 2xy-2y,\nonumber\\
K_{02} &=& 8x+8y+24,\nonumber\\
K_{12} &=& -13xy-8x^2-16y+2y^3-28x-12,\nonumber\\
K_{22} &=& 6y^4+7xy^3-21x^2y-18xy^2-40x^2-166xy+24y^3-56y^2-192x-268y-216,\nonumber\\
K_{32} &=& 13xy^3\!-\!44x^2y\!-\!6x^3\!+\!10y^4\!-\!92x^2\!-\!29xy^2\!+\!36y^3\!-\!285xy\!-\!90y^2\!-\!422y\!-\!338x\!-\!348,\\
K_{42} &=& -12x^2y+4xy^3-13xy^2+4y^4-92x-20x^2-128y-86xy-36y^2+10y^3-96,\nonumber\\
K_{52} &=& -2xy-2x^2+2y^2-2y-6x,\nonumber\\
K_{03} &=& -6y^2+12x+24,\nonumber\\
K_{13} &=& 6xy^2-2xy-12x^2-6y-26x,\nonumber\\
K_{23} &=& -6y^5-35x^2y+30xy^3+31xy^2-62x^2-187xy+82y^3+54y^2-254x-246y-252,\nonumber\\
K_{33} &=& 5x^2y^2-10y^5-10x^3+60xy^2+50xy^3-62x^2y-330xy+138y^3-144x^2 + {}\nonumber\\
       &+& 96y^2-432y-480x-456,\nonumber\\
K_{43} &=& -4y^5+20xy^3-23x^2y+21xy^2-127xy+56y^3-42x^2+36y^2-174y-178x-180,\nonumber\\
K_{53} &=& 2xy^2-2xy-4x^2-6y-10x,\nonumber\\
K_{04} &=& 4x+4y+12,\nonumber\\
\end{eqnarray*}
\begin{eqnarray*}
K_{14} &=& -4xy-4x^2+2y^2-2y-12x,\nonumber\\
K_{24} &=& -12x^2y+4xy^3-13xy^2+4y^4-92x-20x^2-128y-86xy-36y^2+10y^3-96,\nonumber\\
K_{34} &=& -\!30x^2y\!-\!4x^3\!-\!226x\!+\!6y^4\!-\!17xy^2\!+\!9xy^3\!-\!189xy\!-\!62x^2\!+\!24y^3\!-\!54y^2\!-\!274y\!-\!228,\\
K_{44} &=& 3xy^3-9x^2y+2y^4-6xy^2-74xy-20x^2+12y^3-20y^2-96x-124y-108,\nonumber\\
K_{54} &=& -2x^2+2y^3-7xy-10x-16y-12,\nonumber\\
K_{05} &=& -2y,\nonumber\\
K_{15} &=& 2xy-2y,\nonumber\\
K_{25} &=& -2y^4+6xy^2+11xy+20y^2-4x+16y-12,\nonumber\\
K_{35} &=& 2x^2y-4y^4+11xy^2+2x^2+24xy+36y^2+6x+38y,\nonumber\\
K_{45} &=& -2y^4+7xy^2-2x^2+10xy+18y^2-6x+20y,\nonumber\\
K_{55} &=& xy+2y-2y^2+4x+12.\nonumber
\end{eqnarray*}
In conclusion we list here a few polynomials generating by the function (\ref{dks-5})
\begin{eqnarray}
T_{00}&=&1,\quad T_{10}=x,\quad T_{01}=y,\nonumber\\
T_{20}&=&x^2-2y^3+6xy+18y+10x+18,\nonumber\\
T_{11}&=&xy-2y^2+2y+4x+12,\nonumber\\
T_{02}&=&y^2-2x-2y-6\nonumber\\
T_{30}&=&x^3-3y^3x+9x^2y-6y^3+18x^2+45xy+5y+63x+60\nonumber\\
T_{21}&=&x^2y-2y^4+5xy^2+2x^2+12xy+18y^2+6x+20y\nonumber\\
T_{12}&=&xy^2-2x^2+2y^2-3xy-10x-4y-12\nonumber\\
T_{03}&=&y^3-3xy-6x-9y-12.\nonumber
\end{eqnarray}

By the method described in the previous Sections we can obtain the generating function of the
Chebyshev polynomials of the second kind, associated with the algebra $G_2$. Here are a few
polynomials calculated in the work \cite{LU} from the recurrence relations

\begin{eqnarray}
U_{00}&=&1,\quad U_{10}=x,\quad U_{01}=y,\nonumber\\
U_{20}&=&2y+x+x^2-y^3+2xy,\nonumber\\
U_{11}&=&1-y^2+x+xy,\nonumber\\
U_{02}&=&-1-x-y+y^2,\nonumber\\
U_{30}&=&-y^2+2x+3x^2-y^3-2xy^3+y^4+x^3+4xy-2xy^2+4x^2y,\nonumber\\
U_{21}&=&-1-y+2y^2+x^2+y^3-y^4+xy^2+x^2y,\nonumber\\
U_{12}&=&y+y^2-x-x^2-y^3+xy^2,\nonumber\\
U_{03}&=&-y-y^2-x+y^3-2xy.\nonumber
\end{eqnarray}

\vspace{1cm}

\noindent{\bf Acknowledgment}

\vspace{5mm}

\noindent The work was supported by RFBR, grant \No 15-01-03148-а and partially (PPK) by the grant \No 14-01-00341
and the programme ``Mathematical problems of nonlinear dynamics'' of RAS.

\end{document}